\documentclass[10pt,conference]{IEEEtran}

\usepackage[utf8]{inputenc}
\usepackage[T1]{fontenc}
\usepackage{hyperref}
\usepackage{xcolor}
\usepackage{enumitem}
\usepackage{balance}
\usepackage{cite}
\usepackage{booktabs}
\usepackage{todonotes}
\usepackage{graphicx}
\usepackage{balance}

\title{The End of Code Review:\\
Coding Agents Supersede Human Inspection}

\author{
  \IEEEauthorblockN{Martin Monperrus}
  \IEEEauthorblockA{KTH Royal Institute of Technology\\
  Stockholm, Sweden\\
  monperrus@kth.se}
}

\begin{document}

\maketitle

\begin{abstract}
Code review has been the primary quality gate in software development since Fagan formalised code inspection in 1976.
For five decades, having a human examine and comment on a colleague's changes before merge has been a cornerstone practice at organisations of every size.
Coding agents are large language model (LLM)-based autonomous systems capable of reading, writing, testing, and repairing software.
We argue that coding agents have crossed a threshold of capability at which traditional human code review is no longer a necessary component of a software quality pipeline.
Our argument rests on two claims: every stated goal of code review can be served by agents at lower cost and higher throughput; the naive integration in which agents write code and humans remain the mandatory reviewers is a dead end because it neither provides meaningful assurance nor scales with AI-assisted throughput.
\end{abstract}

\section{Introduction}
\label{sec:intro}

Code review is the dominant human quality gate in modern software development.
Since Bacchelli and Bird surveyed its practice at Microsoft~\cite{bacchelli2013expectations} and Sadowski et al.\ documented it at Google~\cite{sadowski2018modern}, the field has understood code review as serving four overlapping goals: finding defects before they reach production, enforcing style and conventions, transferring knowledge between team members, and building shared awareness of the evolving codebase.
No serious software team omits it.

Yet code review carries substantial costs that are often underappreciated.
Developers at large organisations spend between ten and fifteen percent of their working hours reading and commenting on others' code~\cite{sadowski2018modern}.
The review latency between submitting a pull request and receiving actionable feedback routinely stretches over twenty-four hours and can extend to days, imposing a structural drag on continuous-delivery~\cite{hilton2016usage}.
Beyond time, review generates social friction: tone escalations, seniority bias, and the well-documented tendency for first-time contributors to abandon projects after critical feedback~\cite{bosu2016process,murgia2014developers}.

Into this landscape, coding agents have arrived.
LLM-based systems such as Claude Code, Codex, and GitHub Copilot can now read and modify files, execute test suites, interpret compiler output, and iteratively repair failures without human direction~\cite{yang2024sweagent,cognition2024devin,wang2024opendevin,github2024copilotworkspace}.
On SWE-bench (a benchmark of real, unmodified GitHub issues drawn from popular Python libraries), state-of-the-art agents resolve more than eighty percent of tasks end-to-end~\cite{jimenez2024swebench}.
Work on LLM-based code review shows that agents can produce inline defect comments at quality comparable to trained human reviewers~\cite{li2022codereviewer,pornprasit2023automated}.

In this paper, we make a strong claim: \textit{coding agents have reached a capability threshold at which human code review is redundant and should be replaced by agent-driven verification.}
We do not present a new empirical study; instead, we synthesise existing capability evidence, and enumerate the implications for software engineering practice, tooling, and research. Specifically, we argue that agents can satisfy the established goals of review, that the intermediate model in which agents write code but humans remain mandatory reviewers is unstable, and that the economics of mandatory human inspection have already turned negative for routine changes.

To sum up, our contributions are:
\begin{itemize}[noitemsep]
  \item The demonstration that every stated goal of code review---defect detection, style enforcement, knowledge transfer, and team awareness---can be met by coding agents at lower cost and higher throughput than human reviewers.
  \item The argument that combining AI code generation with mandatory human review is not a stable endpoint: it creates the appearance of assurance while turning review capacity into the next delivery bottleneck.
  \item The cost-benefit analysis showing that mandatory human inspection has already flipped; agent reviews are instantaneous, consistent, and auditable, while the marginal defect-detection value of human review shrinks as agent capabilities grow.
\end{itemize}

\begin{figure*}[t]
  \centering
  \includegraphics[width=\textwidth] {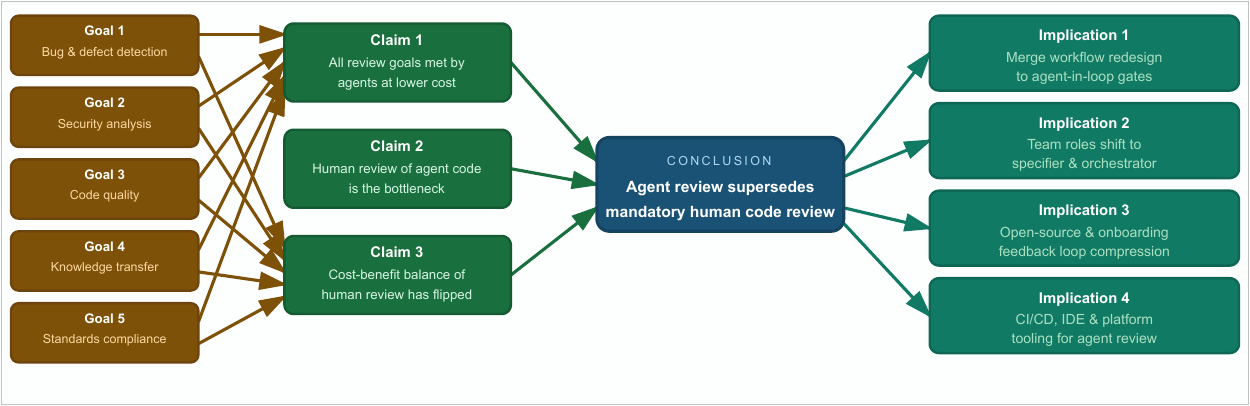}
  \caption{Argument map of the paper. Review goals support three claims that lead to the conclusion, which entails four implications for practice and tooling.}
  \label{fig:argmap}
\end{figure*}

\section{Background}
\label{sec:background}

\subsection{Code Review: History and Practice}

Code review as a formal engineering discipline originates with Fagan's 1976 paper on code inspections at IBM~\cite{fagan1976design}.
Fagan proposed a structured, multi-stage process: planning, overview, preparation, inspection, rework, and follow-up.
The process was presented as expensive, requiring up to three percent of total project effort.

The empirical record on the effectiveness of modern code review is more nuanced than the practice's universal adoption might suggest.
Bacchelli and Bird found that defect detection, while a primary motivation, is not what reviewers most reliably deliver: style corrections, minor improvements, and questions about intent dominate the comment corpus~\cite{bacchelli2013expectations}.
Czerwonka et al.\ reached the pointed conclusion that code reviews at Microsoft ``do not find bugs'' in the sense of catching deep, logic-level defects; their primary value lies in knowledge transfer and maintainability improvement~\cite{czerwonka2015code}.
Knowledge transfer and socialisation emerge consistently across studies as valued outcomes of the review interaction~\cite{bacchelli2013expectations,bosu2016process}.

\subsection{Automated Code Review}

Efforts to automate code review predate large language models.
Early approaches relied on pattern matching, rule-based checkers, and machine-learning classifiers trained on review-comment corpora to predict whether a change would attract a comment and what that comment would say.
These systems automated individual review activities in isolation: comment prediction, priority ranking of changed files, detection of specific defect classes.

The LLM era opened new possibilities.
CodeReviewer~\cite{li2022codereviewer} pre-trains a model on a large corpus of pull-request diffs and associated review comments, achieving competitive performance on comment generation, code-refinement prediction, and severity classification.
LLaMA-Reviewer~\cite{lu2023llama} applies parameter-efficient fine-tuning to adapt a general-purpose LLM for the review-comment generation task, demonstrating that strong review performance is achievable with fine-tuning.
Tufano et al.~\cite{tufano2021using} build an end-to-end pipeline from diff to refined patch, showing that a model can both generate a review comment and produce the corrected code, eliminating one round-trip in the review-revise cycle.
Pornprasit and Tantithamthavorn evaluate these and related systems in industrial settings and find meaningful coverage across defect types~\cite{pornprasit2023automated}.
Tang et al.\ introduce CodeAgent~\cite{tang2024codeagent}, a multi-agent system in which specialised communicative agents collaborate to perform code review, pushing the automation boundary from single-model comment generation toward coordinated agent workflows.
What these systems share is a focus on automating \emph{activities within} a review workflow that remains human-governed; they improve the efficiency of specific steps without questioning whether the overall review gate is necessary.

To our knowledge, this paper is the first to argue for the \emph{complete displacement of mandatory human code review by coding agents}, and to discuss the resulting implications for workflow, tooling and research.

\subsection{Coding Agents}

A coding agent is a system in which a large language model (LLM) is embedded in an \emph{agentic loop}: the model can invoke tools for reading and writing files, executing shell commands, running tests, querying documentation, and then iterate until a goal is achieved or a stopping condition is met.
The LLM provides language understanding, code synthesis, and contextual reasoning; the tool loop provides grounding in the actual state of the software artefact and its execution.

Representative agents include OpenAI Codex~\cite{chen2021evaluating}, an interactive assistant built on top of the GPT line of models; Claude Code~\cite{anthropic2024claude}, an agentic coding assistant that operates directly in the developer's terminal; SWE-agent~\cite{yang2024sweagent}, which introduces an agent-computer interface optimised for repository-scale software engineering tasks; Devin~\cite{cognition2024devin}, a commercial system that autonomously navigates codebases and deploys fixes; and GitHub Copilot~\cite{github2024copilotworkspace}, which integrates agent capabilities directly into the pull-request workflow.

\section{Evidence of Agent Capability}
\label{sec:capability}

To support the claim that coding agents can displace human code review, we survey evidence of agent capability across three dimensions: benchmark performance on software-engineering tasks, review-specific capabilities, and developer productivity with deployed tools.

\subsection{Benchmark Performance}

The most direct evidence for agent capability comes from SWE-bench, a benchmark that evaluates whether a system can resolve real, unmodified issues drawn from popular open-source Python projects~\cite{jimenez2024swebench}.
Unlike synthetic coding challenges, SWE-bench tasks require understanding issue descriptions, navigating multi-file repositories, modifying the correct files, and producing patches that pass the project's existing test suite.
Early results were sobering: GPT-4 with retrieval-based context selection resolved approximately 1.7\,\% of tasks.
SWE-agent, by introducing an agent-computer interface that structures the model's interaction with the file system and shell, raised this figure to approximately 12.5\,\%~\cite{yang2024sweagent}.
By late 2024, the best-performing systems on SWE-bench Verified—a curated subset with verified ground-truth resolutions—exceeded 50\,\%.
By late 2025, top agents on the public leaderboard resolved more than 70\,\% of tasks~\cite{swebench2025leaderboard}.
The improvement trajectory, from under two percent to over seventy percent in roughly two years, is without precedent in the history of automated software engineering tools.

Related evidence comes from adjacent tasks.
For example, Xia et al.\ demonstrate that LLM-based repair approaches substantially outperform earlier generate-and-validate systems like GenProg~\cite{le2012genprog,xia2023automated}, fixing a far larger proportion of benchmark bugs without requiring manually specified test cases or fix templates.
At the competitive programming level, AlphaCode ranked within the top 54\,\% of human participants on Codeforces contests~\cite{li2022competition}. This establishes that LLMs can reason about non-trivial algorithmic problems.

\subsection{Review-Specific Capabilities}

Beyond general software engineering, several strands of work speak specifically to the capabilities that code review requires.
Pornprasit and Tantithamthavorn evaluate LLM-based automated review in industrial settings and find that agents detect the same categories of defect that human reviewers target: correctness errors, security weaknesses, performance inefficiencies, and style violations~\cite{pornprasit2023automated}.
Li et al.\ demonstrate that CodeReviewer produces actionable inline comments at quality that is at least comparable to those of trained human reviewers on a significant fraction of the evaluation set~\cite{li2022codereviewer}.

Agents bring capabilities that human reviewers structurally cannot match.
A human reviewer reads a diff; an agent can simultaneously hold the full file, the complete test suite, the git history of every touched function, and the project's documentation in context.
A human reviewer can identify a problem and describe it in a comment; an agent can identify the problem, generate a fix, apply it, run the tests, and close the review loop without any human scheduling.
A human reviewer has a calendar, a timezone, and finite attention; an agent operates continuously, applying the same scrutiny at 3\,am on a Sunday as during a Monday morning sprint.
These are structural advantages that compound as the scale and velocity of software development increase.

\section{The End of Code Review: The Argument}
\label{sec:argument}

\subsection{Claim 1: Every goal of code review can be served by agents at lower cost and higher throughput.}

Bacchelli and Bird identify four primary goals of code review: defect detection, style and standards enforcement, knowledge transfer, and team awareness~\cite{bacchelli2013expectations}.
We argue that contemporary coding agents satisfy each goal at least as well as human reviewers, and in several dimensions markedly better.

\emph{Defect detection.}
Human reviewers are effective at identifying surface-level issues but unreliable at catching deep logical bugs~\cite{czerwonka2015code}.
\textbf{Agents powered by large language models, by contrast, can perform exhaustive dataflow reasoning, cross-reference the full test suite, and apply learned patterns from millions of open-source repositories.}
Recent work shows that LLM-based review systems produce actionable defect reports comparable in quality to those of trained human reviewers, while operating on every commit without fatigue or time-zone constraints~\cite{pornprasit2023automated,li2022codereviewer}.

\emph{Style and standards enforcement.}
Linters, formatters, and type checkers have long handled syntactic and stylistic concerns automatically.
\textbf{Agents extend this capacity to semantic style: naming consistency, idiomatic API usage, and documentation conventions can all be enforced through LLM-based rewriting, eliminating an entire category of reviewer comment.}

\emph{Security.} 
Security review is precisely the domain where human attention is most overloaded and most consequential.
Pearce et al.\ demonstrate that GitHub Copilot introduces common weakness enumeration (CWE) violations at non-trivial rates~\cite{pearce2022asleep}, yet the same generation capability, when redirected toward \emph{detection}, allows agents to enumerate vulnerability classes far more systematically than a developer performing an ad-hoc review~\cite{khoury2023secure}.
\textbf{AI-based security scanners already outperform many manual reviewers on standard vulnerability benchmarks \cite{lin2025comparing}}.

\emph{Knowledge transfer.}
\textbf{Agents can actively generate on-demand explanations, architectural summaries, and updated documentation at merge time, which is a more reliable and scalable mechanism for propagating knowledge than the incidental commentary of a busy colleague.}

\subsection{Claim 2: The naive integration---AI coding with human review---is a dead end.}

The first response of most organisations to the rise of AI coding tools is as follows: an agent writes the code, a human reviews it.
This arrangement feels conservative and safe.
We argue that it fails on two independent grounds.

\textbf{Human review provides no genuine assurance.}
The traditional rationale for code review rests on a tacit assumption: human developers write code, and human reviewers provide an independent check.
When an agent generates the code, this assumption collapses.
A large language model can produce hundreds of lines of plausible, internally consistent code that contains subtle semantic errors—errors that are invisible without running the full test suite or performing the kind of exhaustive dataflow analysis that only an agent can do systematically.
Human reviewers, reading a diff on a screen, are poorly positioned to catch the errors that AI-generated code may contain.
In practice, reviews of agent-generated code become rubber-stamps: the human approves because the code looks correct, because the tests pass, and because the cognitive cost of genuine scrutiny is prohibitive~\cite{czerwonka2015code}.
The assurance that review is supposed to provide is illusory.

\textbf{Human review does not scale.}
AI coding tools increase developer throughput.
A developer assisted by an agent produces more commits, more pull requests, and more lines of changed code per day than an unassisted developer~\cite{peng2023impact}.
Human review capacity does not scale and 
the result is a bottleneck that grows proportionally with the productivity gain that AI coding delivers.
Organisations that deploy AI coding tools while retaining human-gated review will find that the review queue becomes the binding constraint on their delivery pipeline.
The faster the agents write, the longer the queue grows, and the more review degrades into a formality performed under time pressure.
Retaining human review in this context does not preserve quality; it discards both the productivity benefit of AI coding and the quality benefit of genuine review.

The logically consistent step is to close the loop: an independent agent reviews agent-generated code, with humans intervening only when agents flag uncertainty or a change crosses an explicit risk threshold.

\subsection{Claim 3: The cost-benefit calculation has flipped.}

Code review imposes measurable costs: developers at large organisations spend 10--15\,\% of their working hours reviewing code~\cite{sadowski2018modern}, and review latency introduces delays into continuous-delivery pipelines~\cite{hilton2016usage,shahin2017continuous}.
These costs were historically justified by the defects that review caught.
The justification weakens as agent coverage grows.

Consider the marginal value of a human review comment.
As agents scan every file, every test, and every commit with increasing recall, the set of defects that escape agent review but would be caught by a human reviewer shrinks.
At the same time, the cost of human review in time, in latency, in social friction~\cite{bosu2016process,murgia2014developers}, remains constant.
\textbf{The crossover point, at which the marginal benefit of human review no longer justifies its marginal cost, has already been reached.}
Agent reviews are instantaneous, deterministic, and auditable: they produce structured reports, not informal threaded comments, and they can be re-run at any point in the pipeline.
The competitive advantage of continuous delivery~\cite{shahin2017continuous} makes review latency an increasingly unacceptable tax on software teams, and agents eliminate it entirely.

\section{Implications}
\label{sec:implications}
\subsection{Implications for Software Engineering Practices}
\label{sec:practices}

\textbf{Merge workflow redesign.}
The most immediate implication of agent-based review is the restructuring of the pull-request workflow.
In the current model, a developer opens a pull request and waits, sometimes for days, for a colleague to find time to review it.
We propose replacing this gating step with an \emph{agent-in-the-loop} verification pipeline that runs automatically on every candidate merge.
Merge gates shift from a human approval checkbox to a structured agent sign-off: test coverage thresholds, security scan results, style compliance reports, and reasoning traces are all produced without human scheduling.
Human approval is not eliminated; it is reserved for the decisions that genuinely require it: high-risk changes, novel architecture choices, and regulated code paths where a named human must bear legal accountability.
For the vast majority of commits (incremental features, bug fixes, dependency updates, refactors) agent sign-off is sufficient, and requiring more is waste.

\textbf{Team structure and roles.}
As agent review matures, the role of the professional code reviewer will diminish in its current form.
The time developers spend in synchronous review meetings and asynchronous comment threads will shrink.
This does not mean that developer expertise becomes less important; it means that expertise is redirected.
Developers increasingly become \emph{specifiers} and \emph{orchestrators}: they articulate requirements precisely enough for agents to act on them, evaluate agent-produced artefacts at a higher level of abstraction, and intervene when agent reasoning is uncertain or the stakes are high.
The craft shifts from line-by-line inspection to system-level judgment.
This is a change that many developers will welcome, since surveys consistently show that reviewing code is among the least enjoyable parts of the job~\cite{sadowski2018modern}.

\textbf{Onboarding and knowledge transfer.}
One of the most-cited benefits of code review is its role in propagating knowledge among team members~\cite{bacchelli2013expectations}.
Agents can generate on-demand, context-keyed explanations at merge time that are richer than what a busy senior engineer provides in a rushed comment.
The genuine risk is a reduction in the informal, bidirectional conversations through which tacit knowledge and team culture propagate; organisations will need complementary mechanisms such as pair programming and structured mentorship to preserve this cohesion.

\textbf{Open-source dynamics.}
In open-source projects, maintainer bandwidth is a structural bottleneck: contributions wait weeks for review and projects are abandoned not because patches stop arriving but because maintainers run out of review capacity.
Agent review directly addresses this bottleneck, compressing feedback loops and lowering the barrier for first-time contributors.

\subsection{Implications for Tooling}
\label{sec:tooling}

\textbf{CI/CD integration.}
Agent review is a natural extension of the continuous integration pipeline.
Just as build systems and test runners execute automatically on every push, a review agent can be triggered at the same point, on every commit to a feature branch, or as a mandatory gate before merge.
This reframes review from an asynchronous social activity into a first-class engineering check, analogous to compilation: it runs, it reports, and it either passes or blocks~\cite{shahin2017continuous}.
Agents produce structured reports (JSON, SARIF) that CI dashboards, security platforms, and analytics systems can consume without human mediation.
Review history becomes an engineering artefact, versioned and queryable, rather than a thread of informal comments that ages poorly.

\textbf{IDE and editor integration.}
The latency reduction that agents enable is not confined to the merge pipeline.
An agent embedded in a developer's editor can review changes before they are committed: as the developer writes, the agent reads context, identifies issues, and surfaces them inline.
This collapses the feedback loop from hours or days to seconds.
The interaction model also changes: rather than reading a list of review comments and deciding whether to accept or dismiss each one, the developer converses with the agent in natural language, asking for explanations, requesting alternative implementations, or negotiating tradeoffs.
This is closer to pair programming than to traditional review, and it operates without the social overhead that makes synchronous pairing difficult to sustain at scale.

\textbf{Version control platforms.}
For agent review to function as a first-class activity, platforms such as GitHub and GitLab must extend their identity and permission models to support agent actors.
An agent reviewer needs a cryptographically signed identity, the ability to push records in the platform's audit log, and the capacity to approve, request changes, or block merge in the same way a human reviewer can~\cite{github2024copilotworkspace}.
The pull-request user interface must evolve accordingly: agent-generated summaries, confidence scores, categorised findings, and inline fix suggestions should be presented as structured UI components, not as synthetic comment threads that mimic human behaviour.
Review history in this model becomes machine-readable by design: downstream analytics can operate directly on agent review records (defect trend dashboards, security posture tracking, codebase health scoring) .

\section{Discussion}
\label{sec:discussion}

\subsection{Counter-argument: Hallucination and false negatives.}
LLMs can miss defects that fall outside their training distribution~\cite{khoury2023secure}.
A human reviewer who is unsure about a change will typically say so; an LLM may silently produce an approval.
The primary mitigation is ensemble review: running multiple independent agents, potentially based on different models and prompting strategies, and requiring consensus before sign-off.
A second mitigation is calibrated uncertainty reporting: reviewer agents should be trained and evaluated to abstain (``I don't know'') and emit confidence estimates that track empirical correctness, rather than always issuing a binary approval~\cite{kadavath2022know}.

\subsection{Counter-argument: Security vulnerabilities in agent-generated code.}
Pearce et al.\ demonstrate that GitHub Copilot introduces common weakness enumeration (CWE) violations at non-trivial rates across a range of security-relevant coding scenarios~\cite{pearce2022asleep}.
When the same model family that generates code is also responsible for reviewing it, there is a risk that generative and review blind spots are correlated: the agent may fail to flag the very pattern of vulnerability that its generation tendency produces.
The mitigation is to use cyber-specialized frontier reviewers for security sign-off: recent benchmark-oriented reports show that the latest models can outperform traditional static analyzers and strong baselines on vulnerability-identification tasks~\cite{yildiz2025jitvul,berkeley2025frontiercyberblog}.

\subsection{Counter-argument: Adversarial inputs and prompt injection.}
A sophisticated adversary who can submit a pull request may also craft code that, when read by a reviewing agent, manipulates the agent's reasoning through embedded natural-language instructions.
A maliciously crafted comment, identifier, or string literal could instruct the agent to overlook a vulnerability or to approve a change it should block.
This is an active area of security research with no fully solved defences~\cite{greshake2023indirect}, and it represents a qualitatively new attack surface that does not exist for human reviewers.
Organisations deploying agent review must treat agent reviewer prompt injection as a first-class threat model.

\subsection{Counter-argument: Architectural coherence requires human judgment.}
At the architectural level, reviewers assess whether a change is consistent with the system's long-term design: whether a new abstraction duplicates an existing one, whether a local trade-off accumulates strategic technical debt, or whether an interface decision prevents future extension.
These judgements depend on a mental model of the system's architecture that AI may not yet hold with sufficient fidelity~\cite{bacchelli2013expectations}.

This concern misidentifies the scope of what pull-request review actually delivers.
Architectural coherence is best enforced through design documents, architecture decision records, and dedicated architecture reviews, not through per-commit diff inspection.
Conflating the two overstates what human PR review reliably provides, since empirical studies show that reviewers overwhelmingly focus on surface defects and style rather than strategic architectural validity~\cite{czerwonka2015code,bacchelli2013expectations}.

\subsection{Counter-argument: Ethical accountability requires human judgment.}
At the ethical level, code changes can carry consequences for user privacy, algorithmic fairness, and environmental footprint that require values-based judgement rather than correctness verification.
Agents are calibrated to optimise for technical quality metrics; they are not reliably equipped to detect that a telemetry change violates a user's reasonable privacy expectation or that a ranking modification amplifies demographic bias~\cite{bosu2016process}.
Legal and organisational accountability frameworks assume a named human decision-maker: an automated approval leaves an accountability gap when a change causes harm.


Agents already flag security-sensitive patterns (PII logging, insecure randomness, over-privileged API calls) and their coverage of compliance properties is growing~\cite{pearce2022asleep,khoury2023secure}.
More fundamentally, mandatory PR review is not the appropriate locus for ethical scrutiny of software: this responsibility belongs in requirements engineering and post-deployment monitoring.
Agent review can be governed by the same institutional mechanisms as other automated quality gates, with human escalation reserved for changes that agents flag as uncertain or high-risk.

\section{Conclusion}
\label{sec:conclusion}

Code review has served software engineering for five decades.
From Fagan's formal inspections~\cite{fagan1976design} to the pull-request model that now governs millions of daily commits~\cite{bacchelli2013expectations}, having humans read each other's code before merge has been treated as a mandatory engineering virtue.
We have argued that this assumption no longer holds.

Coding agents have reached a capability level at which every stated goal of code review can be met automatically, faster, and at a scale that human review cannot match.
The transition is already underway for routine changes in industry.
Agents review dependency bumps, refactors, and test additions today, with humans providing oversight only at the margins.
The open question is not whether this shift will occur, but how quickly it will extend from routine changes to the full breadth of software development, and how the profession will manage that extension.

Code review will not disappear overnight.
Its role will refocus to a layer of high-stakes human oversight: architecture decisions with long-lived consequences, security-critical paths in regulated systems, and changes whose correctness depends on requirements that no agent has been given access to.
For everything else, the case for mandatory human review has already weakened to the point where it is difficult to defend on technical grounds.

The end of code review, as we once thought the absolute best practice of modern software engineering, is the beginning of a more productive way to build software.

\balance
\bibliographystyle{IEEEtran}
\bibliography{references}

\begin{thebibliography}{10}
\providecommand{\url}[1]{#1}
\csname url@samestyle\endcsname
\providecommand{\newblock}{\relax}
\providecommand{\bibinfo}[2]{#2}
\providecommand{\BIBentrySTDinterwordspacing}{\spaceskip=0pt\relax}
\providecommand{\BIBentryALTinterwordstretchfactor}{4}
\providecommand{\BIBentryALTinterwordspacing}{\spaceskip=\fontdimen2\font plus
\BIBentryALTinterwordstretchfactor\fontdimen3\font minus \fontdimen4\font\relax}
\providecommand{\BIBforeignlanguage}[2]{{%
\expandafter\ifx\csname l@#1\endcsname\relax
\typeout{** WARNING: IEEEtran.bst: No hyphenation pattern has been}%
\typeout{** loaded for the language `#1'. Using the pattern for}%
\typeout{** the default language instead.}%
\else
\language=\csname l@#1\endcsname
\fi
#2}}
\providecommand{\BIBdecl}{\relax}
\BIBdecl

\bibitem{bacchelli2013expectations}
A.~Bacchelli and C.~Bird, ``Expectations, outcomes, and challenges of modern code review,'' in \emph{Proceedings of the 35th International Conference on Software Engineering (ICSE)}.\hskip 1em plus 0.5em minus 0.4em\relax IEEE, 2013, pp. 712--721.

\bibitem{sadowski2018modern}
C.~Sadowski, E.~Söderberg, L.~Church, M.~Sipko, and A.~Bacchelli, ``Modern code review: a case study at {Google},'' in \emph{Proceedings of the 40th International Conference on Software Engineering: Software Engineering in Practice (ICSE-SEIP)}.\hskip 1em plus 0.5em minus 0.4em\relax ACM, 2018, pp. 181--190.

\bibitem{hilton2016usage}
M.~Hilton, T.~Tunnell, K.~Huang, D.~Marinov, and D.~Dig, ``Usage, costs, and benefits of continuous integration in open-source projects,'' in \emph{Proceedings of the 31st IEEE/ACM International Conference on Automated Software Engineering (ASE)}.\hskip 1em plus 0.5em minus 0.4em\relax ACM, 2016, pp. 426--437.

\bibitem{bosu2016process}
A.~Bosu, M.~Greiler, and C.~Bird, ``Process aspects and social dynamics of contemporary code review,'' in \emph{IEEE Transactions on Software Engineering}, vol.~43, no.~1.\hskip 1em plus 0.5em minus 0.4em\relax IEEE, 2016, pp. 56--75.

\bibitem{murgia2014developers}
A.~Murgia, P.~Tourani, B.~Adams, and M.~Ortu, ``Do developers feel emotions? an exploratory analysis of emotions in software artifacts,'' in \emph{Proceedings of the 11th Working Conference on Mining Software Repositories (MSR)}.\hskip 1em plus 0.5em minus 0.4em\relax ACM, 2014, pp. 262--271.

\bibitem{yang2024sweagent}
J.~Yang, C.~E. Jimenez, A.~Wettig, K.~Lieret, S.~Yao, K.~Narasimhan, and O.~Press, ``{SWE}-agent: Agent-computer interfaces enable automated software engineering,'' \emph{arXiv preprint arXiv:2405.15793}, 2024.

\bibitem{cognition2024devin}
{Cognition AI}, ``Introducing {Devin}, the first {AI} software engineer,'' \emph{Cognition AI Blog}, 2024, \url{https://www.cognition.ai/blog/introducing-devin}.

\bibitem{wang2024opendevin}
X.~Wang, B.~Chen, Y.~Yuan, Y.~Zhang, B.~Li, C.~Qian \emph{et~al.}, ``{OpenDevin}: An open platform for {AI} software developers as generalist agents,'' in \emph{arXiv preprint arXiv:2407.16741}, 2024.

\bibitem{github2024copilotworkspace}
{GitHub}, ``{GitHub Copilot Workspace}: Welcome to the {Copilot-native} developer environment,'' \emph{GitHub Blog}, 2024, \url{https://github.blog/2024-04-29-github-copilot-workspace/}.

\bibitem{jimenez2024swebench}
C.~E. Jimenez, J.~Yang, A.~Wettig, S.~Yao, K.~Pei, O.~Press, and K.~Narasimhan, ``{SWE}-bench: Can language models resolve real-{GitHub} issues?'' \emph{arXiv preprint arXiv:2310.06770}, 2023.

\bibitem{li2022codereviewer}
Z.~Li, S.~Lu, D.~Guo, N.~Duan, S.~Jannu, G.~Jenks, D.~Majumder, J.~Green, N.~Sundaresan, M.~Fu \emph{et~al.}, ``{CodeReviewer}: Pre-training for automating code review activities,'' in \emph{Proceedings of the 30th ACM Joint European Software Engineering Conference and Symposium on the Foundations of Software Engineering (ESEC/FSE)}.\hskip 1em plus 0.5em minus 0.4em\relax ACM, 2022, pp. 1536--1546.

\bibitem{pornprasit2023automated}
C.~Pornprasit and C.~Tantithamthavorn, ``Automated code review in practice,'' in \emph{Proceedings of the 38th IEEE/ACM International Conference on Automated Software Engineering (ASE)}.\hskip 1em plus 0.5em minus 0.4em\relax IEEE, 2023, pp. 394--405.

\bibitem{fagan1976design}
M.~E. Fagan, ``Design and code inspections to reduce errors in program development,'' \emph{IBM Systems Journal}, vol.~15, no.~3, pp. 182--211, 1976.

\bibitem{czerwonka2015code}
J.~Czerwonka, M.~Greiler, and J.~Tilford, ``Code reviews do not find bugs: How the current code review best practice slows us down,'' pp. 27--28, 2015.

\bibitem{lu2023llama}
J.~Lu, L.~Yu, X.~Li, L.~Yang, and C.~Zuo, ``Llama-reviewer: Advancing code review automation with large language models through parameter-efficient fine-tuning,'' in \emph{Proceedings of the 34th IEEE International Symposium on Software Reliability Engineering (ISSRE)}.\hskip 1em plus 0.5em minus 0.4em\relax IEEE, 2023, pp. 647--658.

\bibitem{tufano2021using}
R.~Tufano, S.~Masiero, A.~Mastropaolo, L.~Pascarella, D.~Poshyvanyk, and G.~Bavota, ``Using pre-trained models to boost code review automation.''\hskip 1em plus 0.5em minus 0.4em\relax ACM, 2022, pp. 1--12.

\bibitem{tang2024codeagent}
X.~Tang, K.~Kim, Y.~Song, C.~Lothritz, B.~Li, S.~Ezzini, H.~Tian, J.~Klein, and T.~F. Bissyande, ``{CodeAgent}: Autonomous communicative agents for code review,'' \emph{arXiv preprint arXiv:2402.02172}, 2024.

\bibitem{chen2021evaluating}
M.~Chen, J.~Tworek, H.~Jun, Q.~Yuan, H.~P. d.~O. Pinto, J.~Kaplan, H.~Edwards, Y.~Burda, N.~Joseph, G.~Brockman \emph{et~al.}, ``Evaluating large language models trained on code,'' \emph{arXiv preprint arXiv:2107.03374}, 2021.

\bibitem{anthropic2024claude}
{Anthropic}, ``The {Claude} 3 model family: {Opus, Sonnet, Haiku},'' \emph{Anthropic Technical Report}, 2024.

\bibitem{swebench2025leaderboard}
S.~bench Team, ``{SWE}-bench leaderboard,'' \url{https://www.swebench.com}, 2025.

\bibitem{le2012genprog}
C.~Le~Goues, T.~Nguyen, S.~Forrest, and W.~Weimer, ``A systematic study of automated program repair: Fixing 55 out of 105 bugs for \$8 each,'' pp. 3--13, 2012.

\bibitem{xia2023automated}
C.~S. Xia, Y.~Wei, and L.~Zhang, ``Automated program repair in the era of large pre-trained language models,'' pp. 1482--1494, 2023.

\bibitem{li2022competition}
Y.~Li, D.~Choi, J.~Chung, N.~Kushman, J.~Schrittwieser, R.~Leblond, T.~Eccles, J.~Keeling, F.~Gimeno, A.~Dal~Lago \emph{et~al.}, ``Competition-level code generation with {AlphaCode},'' \emph{Science}, vol. 378, no. 6624, pp. 1092--1097, 2022.

\bibitem{pearce2022asleep}
H.~Pearce, B.~Ahmad, B.~Tan, B.~Dolan-Gavitt, and R.~Karri, ``Asleep at the keyboard? assessing the security of {GitHub Copilot}'s code contributions,'' in \emph{Proceedings of the 43rd IEEE Symposium on Security and Privacy (SP)}.\hskip 1em plus 0.5em minus 0.4em\relax IEEE, 2022, pp. 754--768.

\bibitem{khoury2023secure}
R.~Khoury, A.~R. Avci, J.~Brunelle, and B.~Marc~Camara, ``How secure is code generated by {ChatGPT}?'' 2023.

\bibitem{lin2025comparing}
J.~W. Lin, E.~K. Jones, D.~J. Jasper, E.~J.-s. Ho, A.~Wu, A.~T. Yang, N.~Perry, A.~Zou, M.~Fredrikson, J.~Z. Kolter \emph{et~al.}, ``Comparing ai agents to cybersecurity professionals in real-world penetration testing,'' \emph{arXiv preprint arXiv:2512.09882}, 2025.

\bibitem{peng2023impact}
S.~Peng, E.~Kalliamvakou, P.~Cihon, and M.~Demirer, ``The impact of {AI} on developer productivity: Evidence from {GitHub Copilot},'' \emph{arXiv preprint arXiv:2302.06590}, 2023.

\bibitem{shahin2017continuous}
M.~Shahin, M.~A. Babar, and L.~Zhu, ``Continuous integration, delivery and deployment: A systematic review on approaches, tools, challenges and practices,'' \emph{IEEE Access}, vol.~5, pp. 3909--3943, 2017.

\bibitem{kadavath2022know}
S.~Kadavath, T.~Conerly, A.~Askell, T.~Henighan, D.~Drain, E.~Perez, N.~Schiefer, Z.~Dodds, N.~DasSarma, E.~Tran-Johnson, S.~Johnston, S.~El-Showk, A.~Jones, N.~Elhage, T.~Hume, A.~Chen, Y.~Bai, S.~Bowman, S.~Fort, D.~Ganguli, D.~Hernandez, J.~Jacobson, J.~Kernion, S.~Kravec, L.~Lovitt, K.~Ndousse, C.~Olsson, S.~Ringer, D.~Amodei, T.~B. Brown, J.~Clark, N.~Joseph, B.~Mann, S.~McCandlish, C.~Olah, and J.~Kaplan, ``Language models (mostly) know what they know,'' \emph{arXiv preprint arXiv:2207.05221}, 2022.

\bibitem{yildiz2025jitvul}
A.~Yildiz, S.~G. Teo, Y.~Lou, Y.~Feng, C.~Wang, and D.~M. Divakaran, ``Benchmarking llms and llm-based agents in practical vulnerability detection for code repositories,'' \emph{arXiv preprint arXiv:2503.03586}, 2025.

\bibitem{berkeley2025frontiercyberblog}
{Berkeley Risk and Decisions Initiative}, ``Frontier ai's impact on the cybersecurity landscape (paper summary and blog),'' \url{https://rdi.berkeley.edu/frontier-ai-impact-on-cybersecurity/}, 2025.

\bibitem{greshake2023indirect}
K.~Greshake, S.~Abdelnabi, S.~Mishra, C.~Endres, T.~Holz, and M.~Fritz, ``Not what you've signed up for: Compromising real-world {LLM}-integrated applications with indirect prompt injection,'' \emph{arXiv preprint arXiv:2302.12173}, 2023.

\end{thebibliography}

\end{document}